 \documentclass[final,3p,times,twocolumn,sort&compress]{elsarticle}

 \usepackage{graphicx,amssymb,amsmath,amsthm,lineno}
 \usepackage{dcolumn,multirow}
 \usepackage[center]{subfigure}
 \usepackage{threeparttable}

\biboptions{square}

\journal{Physica A}

\begin{document}

\begin{frontmatter}



\title{Effects of the long-range cohesive forces in binary particle packing dynamics}


\author[ufsa]{Carlos Handrey Araujo Ferraz\corref{cor1}}
\ead{handrey@ufersa.edu.br}
\cortext[cor1]{Corresponding author}
\address[ufsa]{Exact and Natural Sciences Center, Universidade Federal Rural do Semi-\'Arido-UFERSA, PO Box 0137, CEP 59625-900, Mossor\'o, RN, Brazil}

\begin{abstract}
 Studies on random packing of bidispersive particles have shown that such systems can capture the underlying behavior of more complex phenomena found in physics and materials engineering. In industry, bidispersive particles are used to allow the increase of density and fluidity of the formed compounds. The understanding of the dynamics of these processes is therefore of great theoretical and practical interest. In this paper, we perform molecular dynamics (MD) simulations to study the packing process of particles with binary size distribution. Samples with different particle population densities ($p$) as well as different particle size ratios ($\lambda$) have been assessed. The initial positions of five thousand non-overlapping particles are assigned inside a confining rectangular box. After that, the system is allowed to settle under gravity towards the bottom of the box. Both the translational and rotational movements of each particle are considered in the simulations. In order to deal with interacting particles, we take into account both the contact and long-range cohesive forces. The normal viscoelastic force is calculated according to the nonlinear Hertz model, whereas the tangential force is calculated through an accurate nonlinear-spring model. Assuming a molecular approach, we account for the long-range cohesive forces using a Lennard-Jones(LJ)-like potential. The packing processes are studied assuming different long-range interaction strengths. We carry out statistical calculations of the different quantities studied including packing density, radial distribution function and orientation pair correlation function.
\end{abstract}


\end{frontmatter}


\section{Introduction \label{sec:int}}

The random packing of spherical particle has been an interesting tool used to capture the underlying behavior of more complex phenomena for applications in physics and materials engineering such as modeling ideal liquids~\cite{bernal60,bernal64}, granular media~\cite{herrmann95,gonzalez2014}, amorphous materials~\cite{finney76,angell81}, emulsions~\cite{pal2008}, glasses~\cite{lois2009}, jamming~\cite{hern2003}, ceramic compounds~\cite{mcgeary61, hamad90} and sintering processes~\cite{helle85,nair87}. Understanding the structure of random close-packed particles is important because its physical properties may depend on the packing features such as packing density and mean number coordination. The packing density (i.e., the volume ratio occupied by particles to the total aggregate) is affected by the particle size distribution~\cite{kumar2014,kumar2016,ogarko2013}, particle shape~\cite{yuan2018}, and long-range cohesive forces~\cite{gonzalez2014}. In general, random packing structures possess packing densities that increase with increasing width of the size distribution~\cite{sohn1968,schaertl1994,hermes_2010,santos2014}, increasing sphericity, and decreasing long-range cohesive forces. For micro-sized particles, both van der Waals and electrostatic forces play an important role in particle rearrangements as they dominate the dynamical packing process~\cite{visser89,israelachvili92}, forming local particle clusters~\cite{yen91, yang2000, cheng2000, jia12} that can eventuate into large percolation clusters~\cite{handrey2018} depending on the nature of the particles involved.

There have been few earlier experimental and computational studies concerning the micro-sized binary particles packing in which long-range cohesive forces have to be taken into account to describe the adequate behavior of the colliding particles involved in these dynamical processes. Forsyth {\it et al}~\cite{forsyth2001} experimentally investigated the influence of van der Waals forces in hard-sphere packing; however, they did not take into account neither the impacts caused by electrostatic force nor bidispersity. Yen and Chaki~\cite{yen91}, Cheng {\it et al}~\cite{cheng2000} and Yang {\it et al}~\cite{yang2000} each applied a simplified version of the so-called distinct element method~\cite{cundall79} to study the effects of both van der Waals and frictional forces present in hard-sphere packing processes but also did not consider binary particle size distributions in their investigations. More recently, a computational study~\cite{jia12} has considered particle packing dynamics using both Gaussian and bimodal (binary) size distribution in which the van der Waals forces were calculated using the standard Hamaker form~\cite{hamaker1937}, but without including the electrostatic forces between particle pairs or even addressing packing processes upon varying the intensity of the long-range forces. 

The present study has the purpose of addressing the effects of the long-range cohesive forces on packing dynamics with binary particle size distribution by making use of a molecular approach of these forces through a modified version of the Lennard-Jones (LJ) potential, completing thus prior works~\cite{jia12,handrey2019} on particle packing with non-Gaussian distributions. The relationship between packing observables, such as density and radial distribution function (RDF), and dominating long-range forces might be useful for better understanding the interconnection between microscopic and macroscopic properties of the particle packing. 

More specifically, for large particulate systems, such molecular approach can be guaranteed provided it is reminded that when two microspheres (with radius $r$) are separated by a certain distance $d >> r$, the effective potential ($\Phi$) is analogous to that between two molecules, i.e., falling off as $\Phi(d)\varpropto -1/d^6$~\cite{israelachvili92, lyklema91}. Assuming the validity of this modified LJ approximation, we are able to account for the long-range forces involved in the packing process in a simple way. The main advantage of this assumption is to embed both electrostatic (repulsive electronic clouds) and van der Waals forces in the simulations by using few control parameters. This approximation has therefore allowed us to study a variety of different packing cases by considering LJ particles with different potential well depths, which play a dominant role in the strength of these long-range forces.

In this paper, we perform molecular dynamics (MD) simulations to study the packing process of spheres using binary particle size distribution. Both the translational and rotational movements of each particle are considered in the simulations.  We take into account both the contact and long-range cohesive forces. The contact forces result from the deformation of the colliding particles, which can be decomposed into two main types: the normal viscoelastic force and the tangential force. The normal viscoelastic force is calculated according to the nonlinear Hertz model~\cite{popov2010, brilliantov1996}, whereas the tangential force is calculated through a nonlinear-spring model that is derived from the Mindlin--Deresiewicz theory~\cite{mindlin1953}. Based on the molecular framework assumption, we account for the long-range forces using a modified LJ potential. The packing processes were studied by applying different long-range interaction strengths. We performed statistical calculations of the different quantities studied including kinetic energy, packing density and mean coordination number as the system evolved over time. Furthermore, we calculated the radial distribution and orientation pair correlation functions of the formed packing structures.

The content of the manuscript is organized as follows. In section~\ref{sec:mms}, we describe in detail, the model and MD simulations. In section \ref{sec:r}, we  present and discuss the results. Lastly, in section \ref{sec:c}, we draw the conclusions. 

\section{ Model and Molecular Dynamics Simulation \label{sec:mms}}

\begin{figure*}[!t]
\centering
\begin{minipage}[t]{1.0\linewidth}
\centering
\subfigure[]{\label{fig:01a}\includegraphics[scale=0.36, angle=0]{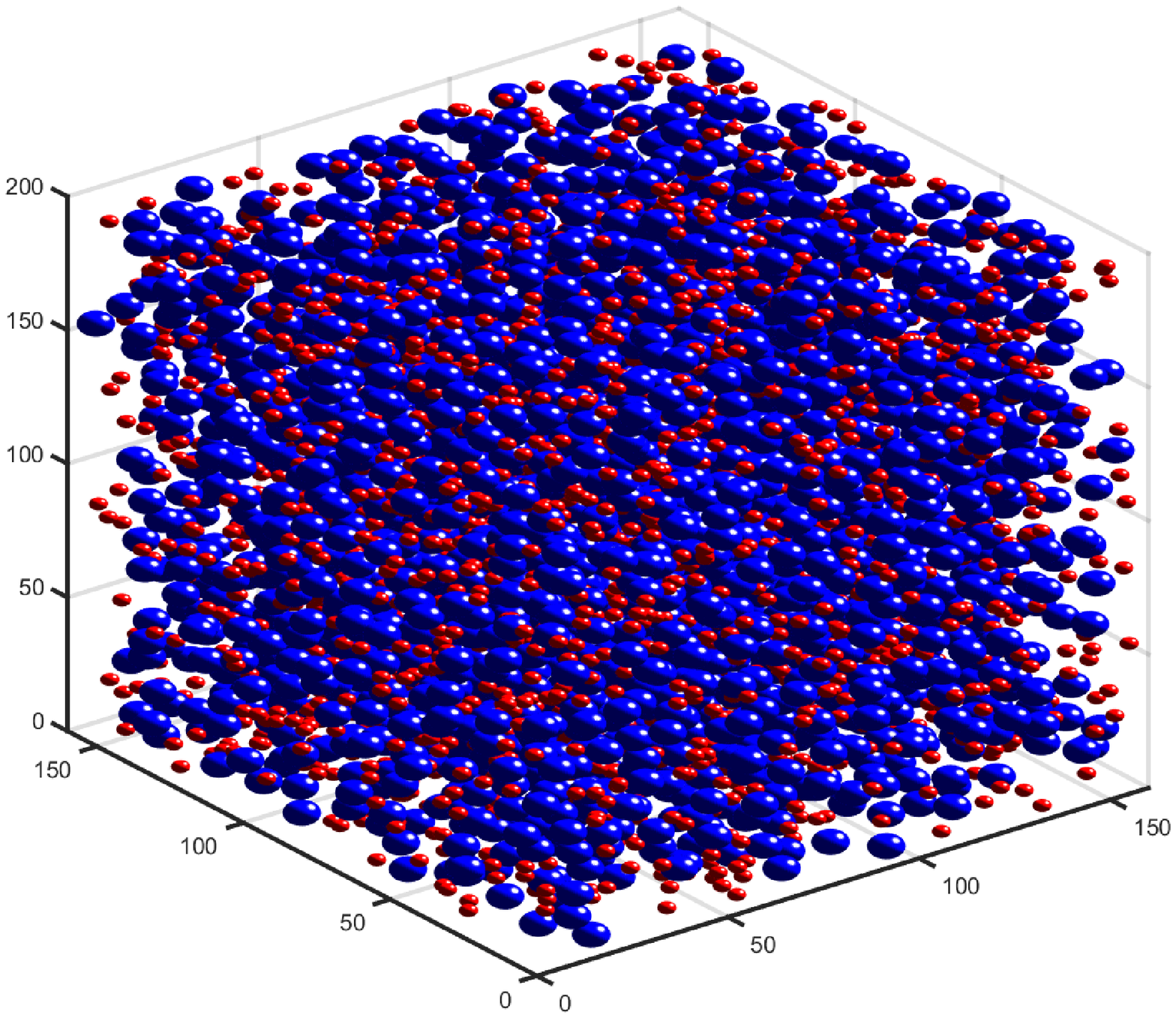}}
\subfigure[]{\label{fig:01b}\includegraphics[scale=0.30, angle=0]{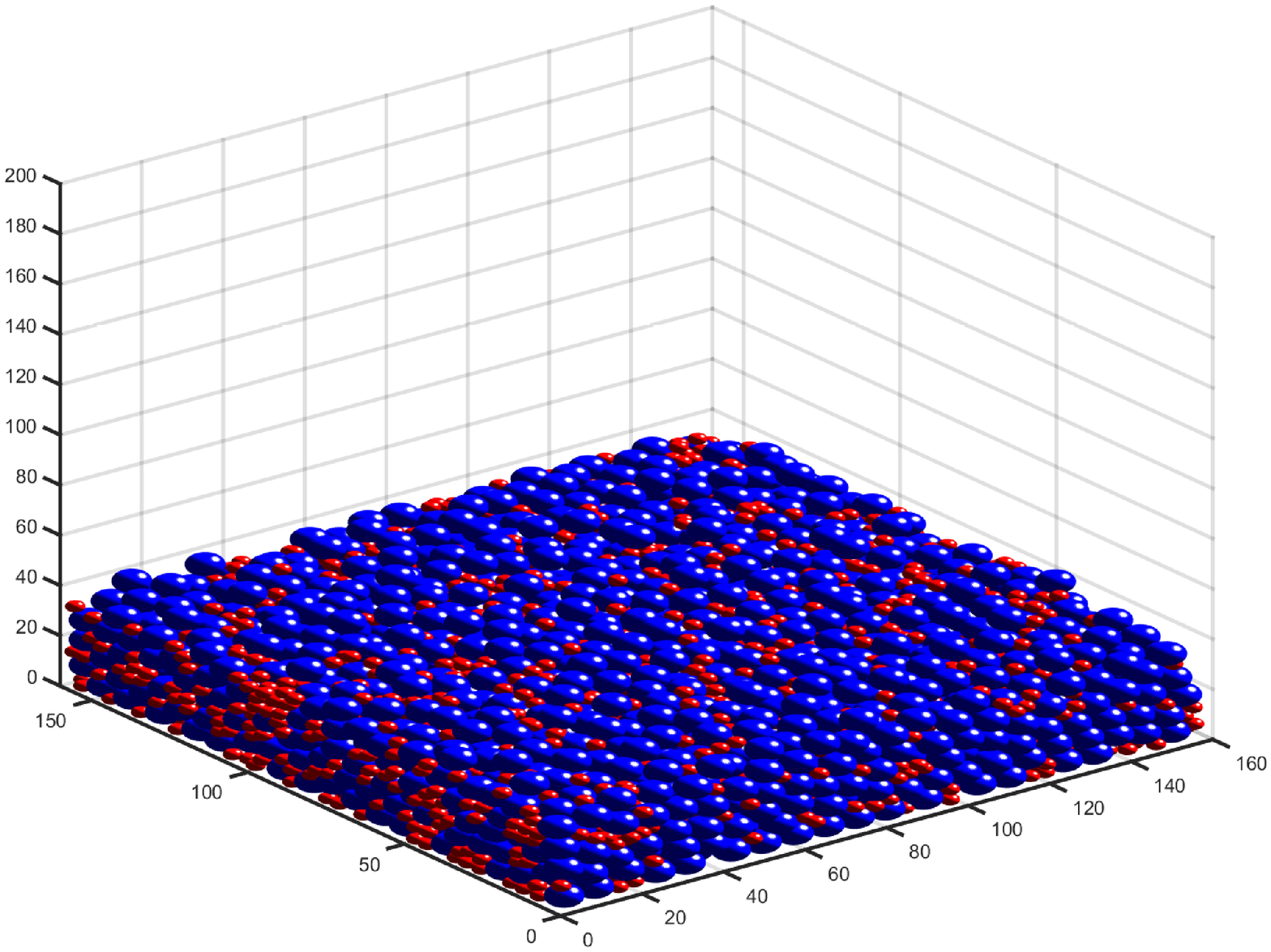}}
\end{minipage}
\caption{Snapshots of a typical packing process with binary distribution inside a $(160 \times 160 \times 200) \, \mu$m box at the instants (a) $t=0$ ms and (b) $t=10$ ms.  The parameters used in this simulation are given in Table \ref{table:01} with a population density $p=0.50$ and an interaction strength of $\varepsilon=10.0\, \mu$J. Large particles ($a=4.0\, \mu$m) are in blue and small particles ($b=2.0\, \mu$m) are in red. }\label{fig:01}
\end{figure*}

 The equations of motion of an $i$-th particle of mass $m_{i}$ and radius $R_{i}$ are:

\begin{equation} \label{eq:1}
{m_i}\frac{{{d^2}{{\vec r}_i}}}{{d{t^2}}} = \sum\limits_j {(\vec F_{ij}^n + \vec F_{ij}^t + \vec F_{ij}^{LJ}) + {m_i}\vec g }
\end{equation}
and 
\begin{equation} \label{eq:2}
 {I_i}\frac{{d{{\vec \omega }_i}}}{{dt}} = \sum\limits_j {{R_i}\,{{\hat n}_{ij}} \times \vec F_{ij}^t}-\gamma _{r} \,R_i|\vec F_{ij}^n|\,{\vec \omega }_i,
\end{equation}
where $\vec{r}_i$ is the position, $\vec{\omega}_i$ is the angular velocity, $\hat{n}_{ij}$ is the unity vector in the direction $j\rightarrow i$, $\gamma _{r}$ is the rolling friction coefficient and $I_{i}=(2/5) \,m_{i}R_{i}^{2}$ is the moment of inertia of the particle. 

In the above equations, $\vec F_{ij}^n$ is the normal viscoelastic force, $\vec F_{ij}^t$ is the tangential friction force, $\vec F_{ij}^{LJ}$ is the LJ force between the $i$- and $j$-th particle, and $\vec g$ is the gravity acceleration. The normal viscoelastic force $\vec F_{ij}^n$ is derived from the nonlinear Hertz theory, and it can written as
\begin{equation} \label{eq:3}
 \vec F_{ij}^n = [\frac{2}{3}E\sqrt {\bar R}\, \delta _{n}^{3/2}-{\gamma _n}E\sqrt {\bar R} \sqrt {{\delta _{n}}} ({{\vec v}_{ij}} \cdot {{\hat n}_{ij}})]{{\hat n}_{ij}},
\end{equation}
where $E$ is the elastic modulus of the two particles, $\bar R=R_{i}R_{j}/(R_{i}+R_{j})$ is the effective radius, $\delta _{n}$ is the deformation which is expressed by
\begin{equation}\label{eq:4}
	\delta _{n}=(R_{i}+R_{j})-(|\vec{r}_{i}(t)-\vec{r}_{j}(t)|),
\end{equation}
 $\vec{v}_{ij}$ is the relative velocity between $i$- and $j$-th particle, and $\gamma _{n}$ is the normal damping  coefficient. 

\begin{table}[!b]
\centering
\small
\begin{threeparttable}
\caption{\label{table:01} Parameters used in the simulations.}
\begin{tabular}{lc}
\hline \hline \\
  Parameter\tnote{a} &  Value \\ \hline \\
  Number of particles ($N$) & 5000 \\
  Particle size scale ($r$) & $1.0<r<10.0\,\mu$m \\
  Particle density ($\rho$) & $2500/\pi\;$ kg/m$^{3}$ \\
	Minimum potential energy ($\varepsilon$) & $0-35.0\,\mu$J\\
  Young's modulus ($Y$) &  $10^{8}\,$ N/m$^{2}$ \\
	Normal damping coefficient ($\gamma_{n}$) &  $5.0\times 10^{-5}$ s \\
  Poisson's ratio ($\xi$) & 0.30 \\
  Tangential damping coefficient ($\gamma_{t}$) &  0.30 \\
	Rolling friction coefficient ($\gamma_{r}$) & 0.002 \\
\hline \hline
\end{tabular}
\begin{tablenotes}
      \item[a]{It is assumed that both particles and walls have the same physical parameters.}
    \end{tablenotes}
\end{threeparttable}
\end{table}

The tangential friction force $\vec F_{ij}^t$ is calculated according to the Mindlin--Deresiewicz theory as
\begin{equation} \label{eq:5}
\vec F_{ij}^t = {\gamma _t}|\vec F_{ij}^n|\left[ {1 - {{\left( {1 - \frac{{|{\delta _{t}}|}}{{|{\delta _{\max }}|}}} \right)}^{3/2}}} \right]{{\hat t}_{ij}},
\end{equation}
where  $\gamma _t$ is the friction coefficient, ${\hat t}_{ij}$ is the unit vector perpendicular to ${\hat n}_{ij}$, $\delta _{t}$ is the tangential displacement which is determined as
\begin{equation} \label{eq:6}
{\delta _{t}} = \int\limits_{{0}}^{t_{c}} {({{\vec v}_{ij}} \cdot {\kern 1pt} {\kern 1pt} {\kern 1pt} } {{\hat t}_{ij}} + {R_i}\,{{\hat n}_{ij}} \times {{\vec \omega }_i} + {R_j}\,{{\hat n}_{ij}} \times {{\vec \omega }_j})dt,
\end{equation}
where the above integral is calculated during the contact time $t_{c}$ between the particles~\cite{schwager1998, landau1965}. The $\delta _{\max }$ is the maximum tangential displacement and in the condition that $\delta _{t}>|\delta _{\max}|$, the sliding friction takes place between the particles. In Eqs.~\ref{eq:3} and \ref{eq:5}, $E$ and $\delta _{\max }$ are given, respectively, by
\begin{equation}\label{eq:7}
	E=Y/(1-\xi^2)
\end{equation}
and 
\begin{equation} \label{eq:8}
	{\delta _{\max }} = {\gamma _t}\frac{{2 - \xi }}{{2(1 - \xi )}}{\delta _{n}},
\end{equation}
being $Y$ the Young's modulus and $\xi$ the Poisson' ratio. 

The relative loss of energy during the collision of two viscoelastic spheres can be measure by evaluating the coefficient of restitution defined by
\begin{equation}\label{eq:8a}
e =  - \frac{{\dot \delta _n (t_c )}}{{\dot \delta _n (0)}}.
\end{equation}
In above equation the impact velocity $v \equiv\dot \delta _n (0)$ is given by
\begin{equation}\label{eq:8b}
v = \frac{{[\vec v_i (0) - \vec v_j (0)] \cdot [\vec r_i (0) - \vec r_j (0)]}}{{R_i  + R_j }}.
\end{equation}

\begin{figure}[!t]
 \centering
 \includegraphics[scale=0.34]{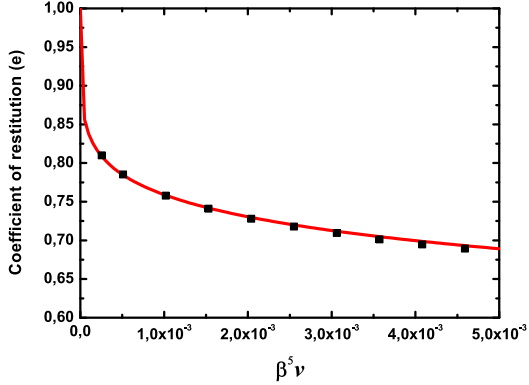}
 \caption{Plot of the Eq.~\ref{eq:8c} for the coefficient normal of restitution $e$ as a function of the impact velocity, considering up to 20 first terms in the series expansion of $e$ (red line). Black points are numerical outcomes coming from collision simulations between two spheres of radii equal to $4.0\,\mu$m by considering different impact velocities.} \label{fig:13}
\end{figure}

Solving analytically the normal part of the Eq.~\ref{eq:1} for the trajectory $\delta _n(t)$ of a pair of colliding particles, it can be shown~\cite{schwager2008} that the coefficient of restitution is described by a power series of $v^{1/10}$:  
\begin{align}	\label{eq:8c}
e(v) &= 1+\sum\limits_{k = 1}^\infty a_k (\beta^{5} v )^{k/10}   \\   
  &= 1 - 1.153\beta v^{1/5}  + 0.798\beta ^2 v^{2/5}  + 0.267\beta ^{5/2} v^{1/2}   \nonumber \\ 
	&- 0.523\beta ^3 v^{3/5}  - 0.461\beta ^{7/2} v^{7/10}  + 0.349\beta ^4 v^{4/5}  +  \ldots \nonumber,
\end{align}
where $\beta  = (2/3)^{-3/5} \gamma _n \,E^{2/5} \bar R^{1/5}$, and a more complete set of coefficients $\{ a_{k}\}$ are found in Ref.~\cite{schwager2008}. In Fig.~\ref{fig:13}, we plot the coefficient normal of restitution $e(v)$ as a function of the impact velocity, considering up to 20 first terms in the series of the Eq.~\ref{eq:8c}. For a collision between two spheres of radii equal to $4.0\,\mu$m, $\beta \simeq 0.12$ s$^{1/5}$/m$^{1/5}$.  

The LJ force between the particles $i$ and $j$ can be evaluated as
\begin{equation} \label{eq:10}
\vec F_{ij}^{LJ} = \frac{{24\varepsilon }}{\sigma }\left[ {2{{\left( {\frac{\sigma }{{{r_{ij}}}}} \right)}^{13}} - {{\left( {\frac{\sigma }{{{r_{ij}}}}} \right)}^7}} \right]{{\hat n}_{ij}},
\end{equation} 
where $r_{ij}$ is the distance between the particles, $\varepsilon$ is the well depth of the LJ potential, which rules the strength of the interaction, and $\sigma=2^{-1/6}(R_{i}+R_{j})$ defines the hard core of the potential. Here, it is important to say that the LJ force is only activated when $r_{ij}>R_{i}+R_{j}$. For $r_{ij}\leq R_{i}+R_{j}$, the contact forces given by Eqs.~\ref{eq:3} and \ref{eq:5} take over control of the particles' driving.  Besides that, we have also used a cutoff at $r_{ij}=3\,(R_{i}+R_{j})$ for saving time during the simulations. 

Due to its effectiveness and symplectic feature, a velocity Verlet algorithm~\cite{rapaport95} was used to integrate the Eqs.~(\ref{eq:1}) and (\ref{eq:2}). The frictional forces helped to ensure stable simulations when taking a time-step $\delta t = 10^{-6}$ s. The average CPU time per particle required to update the phase-space coordinates of the system was around $0.75 \thinspace \mu$s on one 3.70 GHz Intel microprocessor.

\section{\label{sec:r} Results and Discussion}

In this work, the particle packing processes were investigated using binary particle size distributions with different population density $p$ for smaller particles, i.e., the formed samples are composed by a density $p$ of smaller particle and a density $q=1-p$ of bigger ones. In addition, we have also considered different long-range interaction strengths $\varepsilon$. The initial positions, as well as the radii of $5000$ non-overlapping particles were assigned inside a confining $(160 \times 160 \times 200)\,\mu$m box by using a random number generator~\cite{recipes96}. As a way to get around the cumbersome problem of inserting non-overlapping particles inside the box, the bigger particles were randomly put into the box before the smaller ones. Moreover, in order to avoid the complicating effects of the pouring rate, the particles were suspended along the box at the beginning of the simulation. After that, the particles were pulled down by gravity and started to collide each other. Here, no periodic boundary conditions were assumed and, hence, the particle-wall interactions had also to be taken into account. 

The binary distribution is represented by
\begin{equation} \label{eq:11}
f(r) = (1 - p)\delta (r - a) + p\delta (r - b),
\end{equation}
where, as explained above, $p$ is the population density for smaller particles of radius $a$ and $\delta (r-x)$ is the Dirac delta function defined by
\begin{equation}\label{eq:11b}
\delta (r - x) = \frac{1}{{2\pi }}\int\limits_{ - \infty }^\infty  {e^{i\omega (r - x)} d\omega }. 
\end{equation}
Therefore, the binary distribution consists of particles with two distinct radii. The shape of this distribution is determined by the size ratio $\lambda=b/a$, which is the ratio of radii of small particles to large particles, and by the population density (number ratio) $p=f(b)/(f(a)+f(b))$. Note that $\int_a^b {f(r)dr = 1}$. In general, by defining the moments of $r$ as $<r^n >  = \int {r^n } f(r) dr$, it is easy to obtain the mean radius
\begin{equation}\label{eq:12}
	 < r >  = (1 - p)a + pb,
\end{equation}
 and the standard deviation
\begin{equation}\label{eq:13}
\sqrt { <\Delta r^2 > }  = |a - b|(p(1 - p))^{1/2}. 
\end{equation}
Where the relative width, or polydispersity $\Delta r^*$, is then given by
\begin{equation}\label{eq:14}
\Delta r^* = \frac{{\sqrt {<\Delta r^2>} }}{{ < r > }}.
\end{equation}
 Surprisingly, it has been shown~\cite{sohn1968,schaertl1994,hermes_2010,santos2014} that as the polydispersity increases, the packing density also increases as smaller particles can pack more efficiently either by wrapping around larger particles or by fitting into the voids created between neighboring large particles.

\begin{figure*}[!t]
\centering
\begin{minipage}[t]{1.0\linewidth}
\centering
\subfigure[Case $p=0.20$]{\label{fig:02a}\includegraphics[scale=0.36, angle=0]{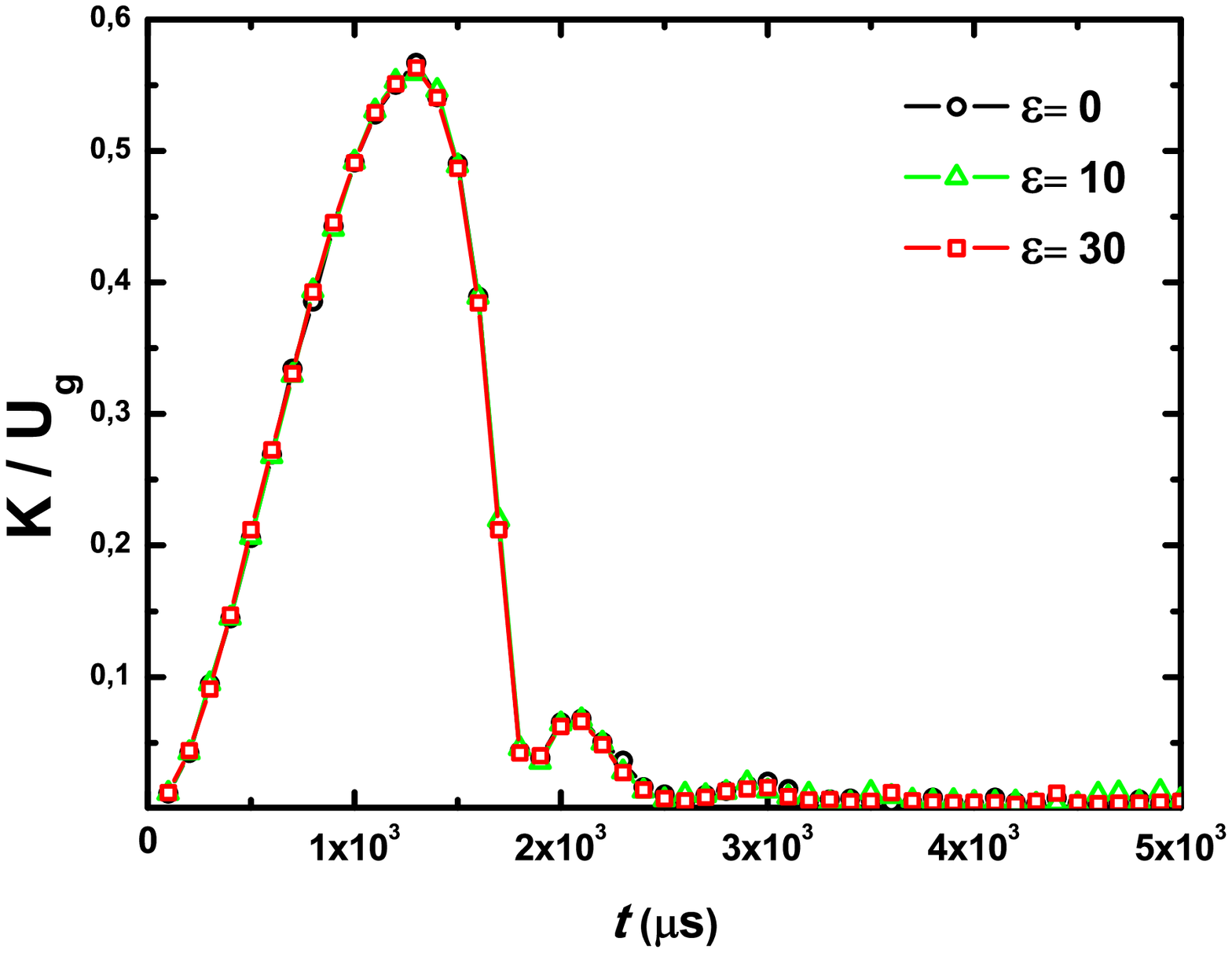}}
\hspace{1.0cm}
\subfigure[Case $p=0.70$]{\label{fig:02b}\includegraphics[scale=0.36, angle=0]{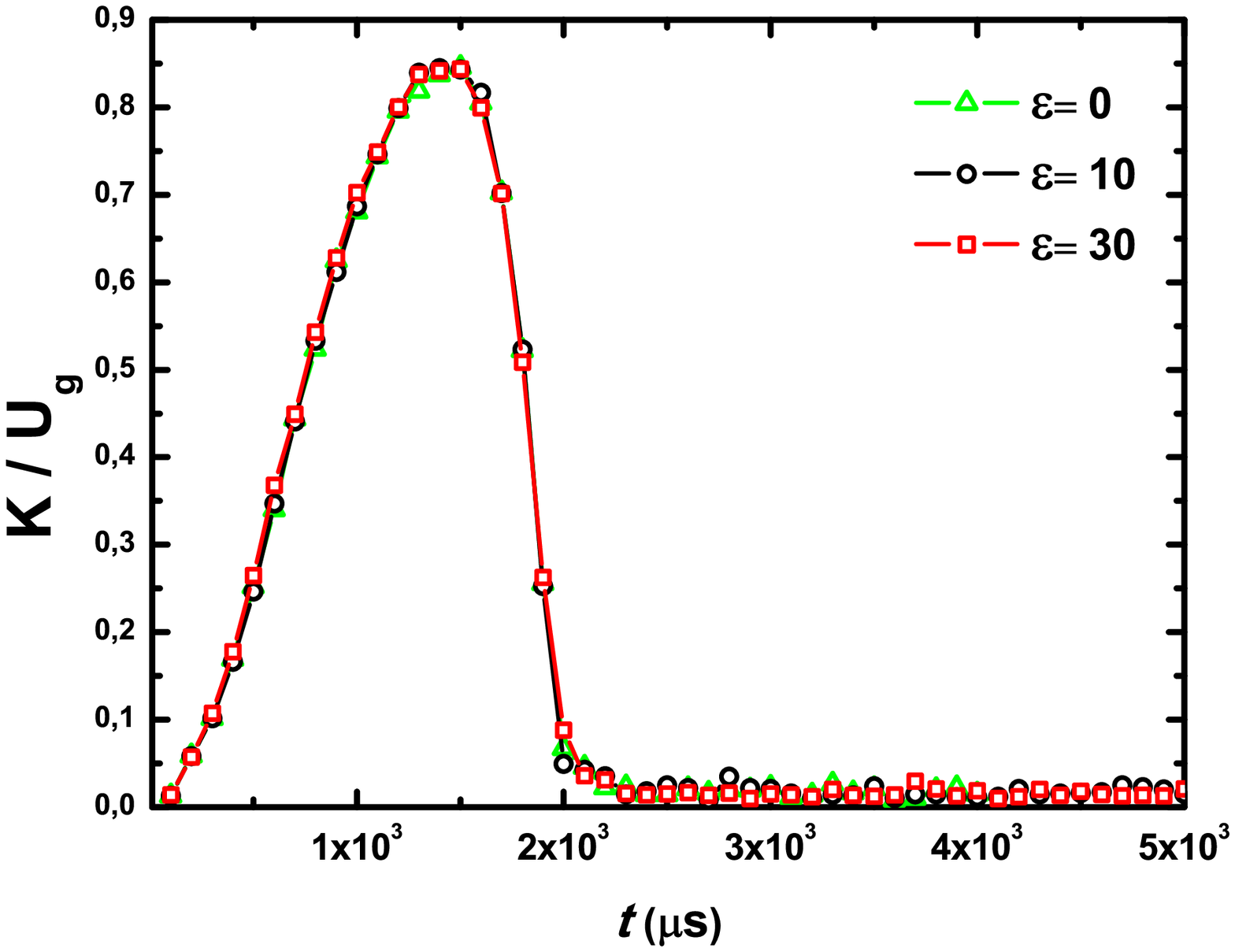}} 
\end{minipage}
\caption{Plot of the average ratio of the kinetic energy ($K$) to the potential energy ($U_{g}$) per particle for a particle binary distribution ($a=4.0\, \mu$m, $b=2.0\, \mu$m) as a function of time, considering three different $\varepsilon$ values. Both the cases $p=0.20$ (a) and $p=0.70$ (b) are shown. The steady state is reached more rapidly for the case $p=0.70$ than for the case $p=0.20$. In this state the force ratios defined by Eq.~\ref{eq:14b} are: $\zeta=0$ ($\varepsilon=0\,\mu$J); $\zeta=0.04$ ($\varepsilon=10\,\mu$J); and $\zeta=0.12$ ($\varepsilon=30\,\mu$J) for the case $p=0.20$. While for the case $p=0.70$: $\zeta=0$ ($\varepsilon=0\,\mu$J); $\zeta=0.09$ ($\varepsilon=10\,\mu$J); and $\zeta=0.27$ ($\varepsilon=30\,\mu$J).}\label{fig:02}
\end{figure*} 

The packing process is depicted in Fig.~\ref{fig:01} for particles with binary distribution by using a population density of $p=0.50$ and a long-range interaction strength $\varepsilon=10.0\, \mu$J. Snapshots at the instants $t=0.0$ ms and $t=10.0$ ms are shown in this figure. The small particles ($b=2.0\, \mu$m) are rendered in red, while large ones ($a=4.0\, \mu$m) are rendered in blue. The parameters used in the simulations are given in Table \ref{table:01}. We performed statistical calculations of different quantities such as packing density, mean coordination number and  kinetic energy as the system evolved over time.  To determine the average value of these quantities and estimate their statistical error, we averaged over $10$ independent realizations.  

Furthermore, as earlier studies~\cite{Desmond2014,seidler2000} have shown particles can form successive layers against the confining walls, and such confinement induces changes in structural quantities near the walls, with a decay toward the bulk values characterized by length scales comparable to the small particle diameter ($\sim2b$) for the binary particle packing. The present work is in line with such studies since all packing quantities were calculated inside a smaller virtual box with virtual walls distant at least $5\, \mu$m from the actual walls of the confining box. In order to check the wall-induced effects on the packing observables, we have also imposed periodic boundary conditions in the lateral walls of the box and redone the simulations for some cases. However, it was seen that flat walls had very little influence on the final density (about $1\%$). It is also worth mentioning that the increase of the particles pile height ($h$) may increase the final packing density ($\phi$) as upper layers would exert more load on lower layers resulting in more dense packings. However, such complicating effect is outside of the scope here, and an adequate investigation of how $h$ variations change the final value of $\phi$ would involve much more simulations taking into account also different box sizes. 

In Fig.~\ref{fig:02} is shown the time evolution of the average ratio of the kinetic energy ($K$) to the potential energy ($U_{g}$) in the cases $p=0.20$ and $p=0.70$ for three different $\varepsilon$ values. Similar energy curves were found for other remaining cases. From this figure, one can see that the system relaxation was already achieved around 2.0 ms for all $\varepsilon$ values considered. However, the system steady state was reached after 3.0 ms for the case $p=0.20$ and after 2.0ms for the case $p=0.70$. The longer equilibration time for the case $p=0.20$ is mainly due to the greater number of large particles present, which possess more inertia, bouncing more times after hitting the bottom of the box until they have their movements halted by the dissipative forces. That also explains the lower peak seen in the curves of energy ratio for this case.

Besides that, in the caption of the Fig.~\ref{fig:02} the number $\zeta$ of the average ratio between relevant forces for a single particle is also given for each case.  The number $\zeta$ is defined by

\begin{equation}\label{eq:14b}	
\zeta  = \frac{{ < F^{LJ} > }}{{ < mg > }}.
\end{equation}   

\begin{figure*}[!t]
\centering
\begin{minipage}[t]{1.0\linewidth}
\centering
\subfigure[$p=0.20$]{\label{fig:03a}\includegraphics[scale=0.34, angle=0]{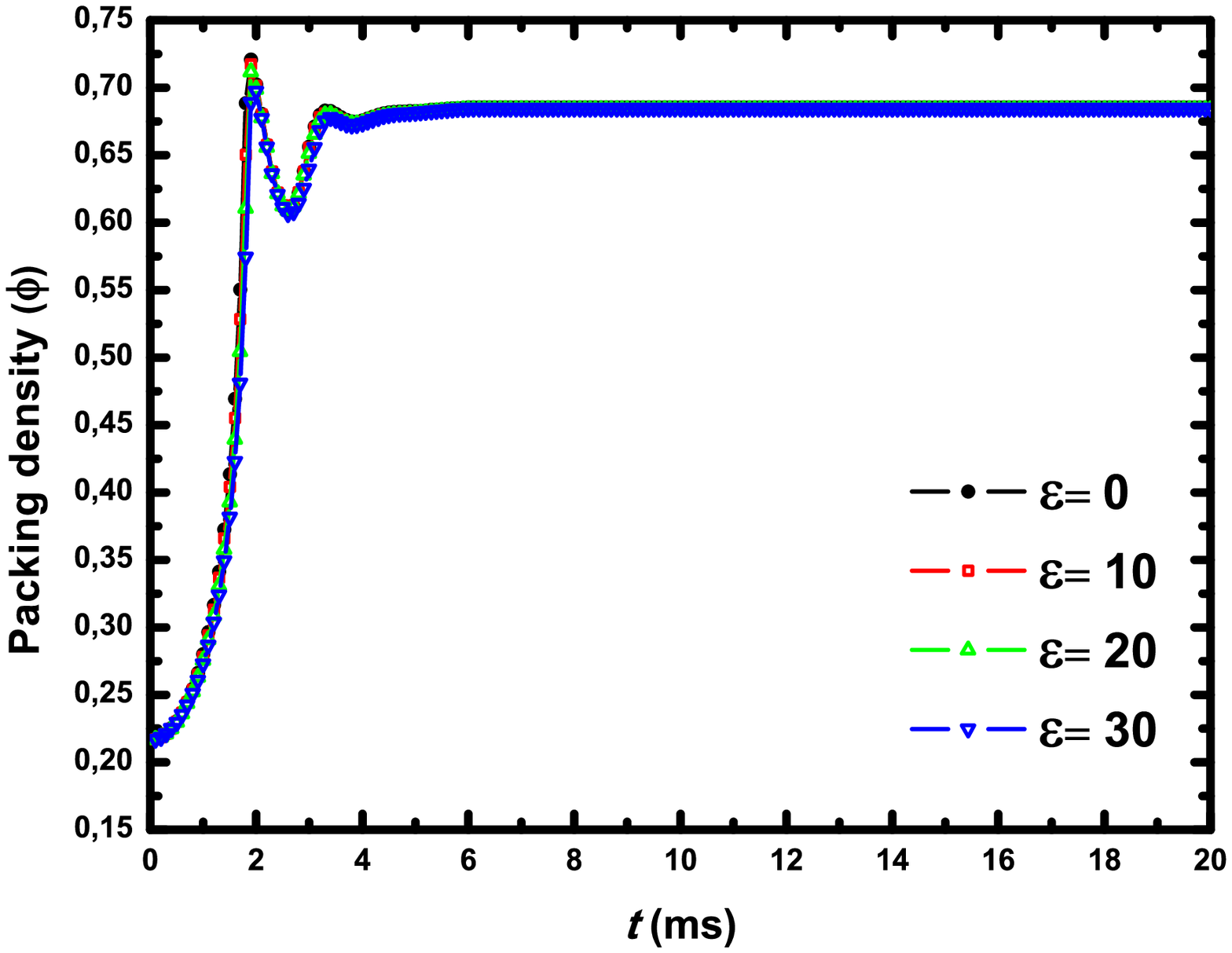}}
\qquad
\subfigure[$p=0.50$]{\label{fig:03b}\includegraphics[scale=0.34, angle=0]{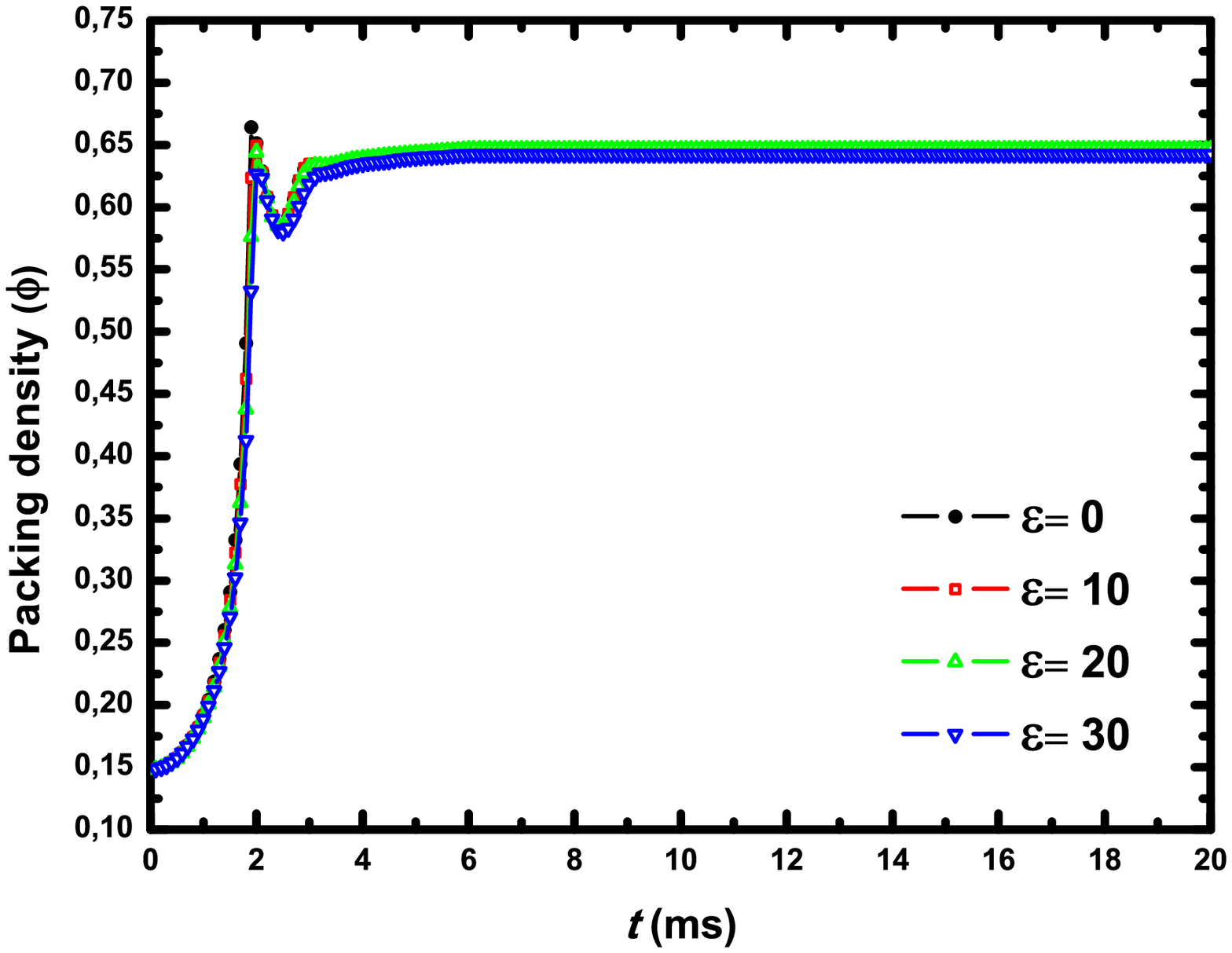}} 
\subfigure[$p=0.70$]{\label{fig:03c}\includegraphics[scale=0.34, angle=0]{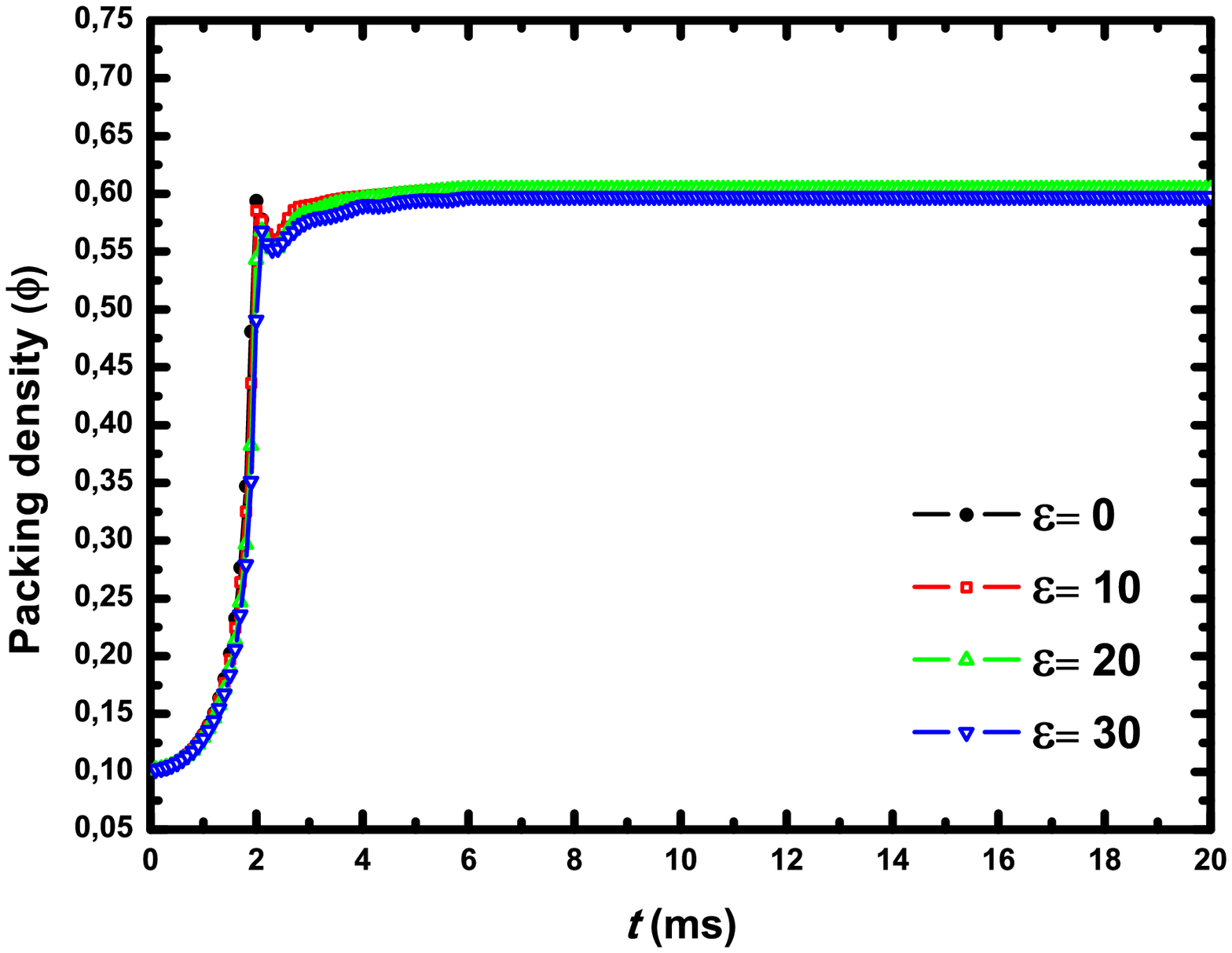}}
\qquad
\subfigure[$p=1.0$]{\label{fig:03d}\includegraphics[scale=0.34, angle=0]{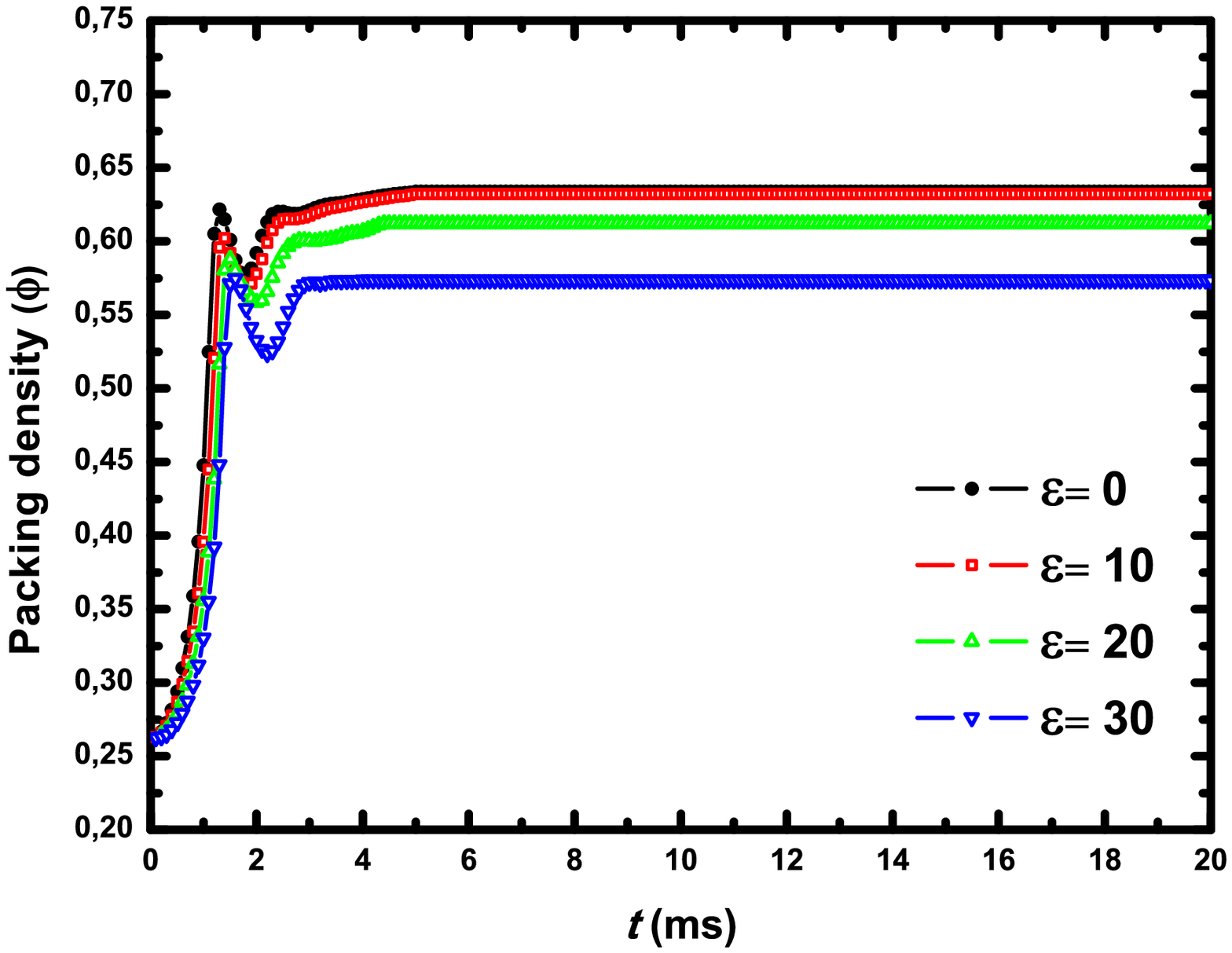}}
\end{minipage}
\caption{Plot of the packing densities $\phi$ for different population probabilities $p$ as a function of time ($a=4.0\, \mu$m, $b=2.0\, \mu$m). Several interaction strengths $\varepsilon$ are considered for each case. Data points are averages over 10 independent realizations.}\label{fig:03}
\end{figure*}

\begin{figure*}[t]
\centering
\begin{minipage} [t]{0.49\linewidth}
\centering
\includegraphics*[scale=0.36]{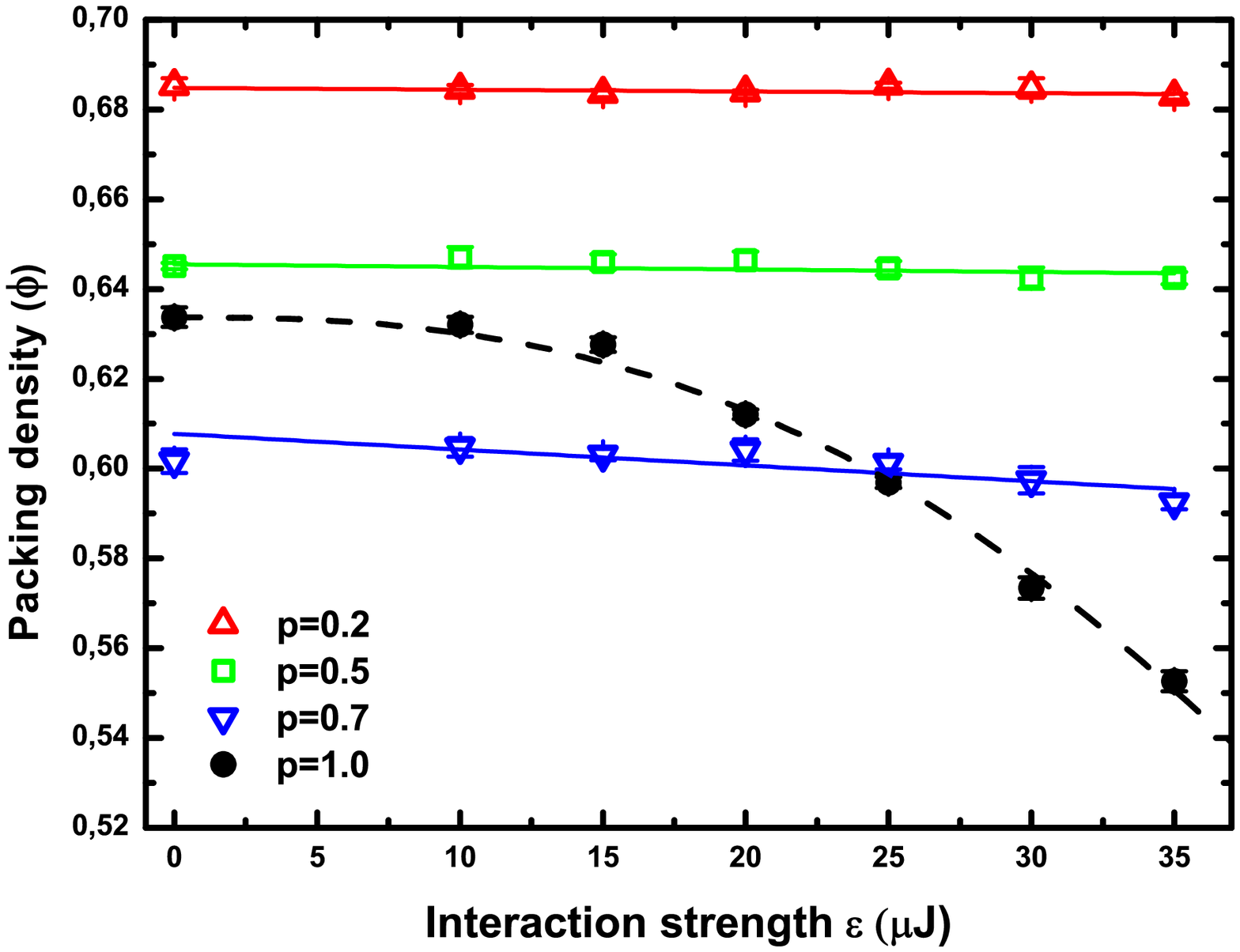}
 \caption{The ultimate $\phi$ values from Fig.~\ref{fig:03} as a function of the interaction strength $\varepsilon$ for all cases $p$ considered ($a=4.0\, \mu$m, $b=2.0\, \mu$m). The solid straight lines are linear fits to data points, whereas the dashed line is the best non-linear fit according to Eq~\ref{eq:16}. Error bars were calculated by averaging 10 independent runs and are not larger than the size of the plotting symbols.} \label{fig:04}
\end{minipage}\hfill
\begin{minipage}[t]{0.49\linewidth}
\centering
\includegraphics[scale=0.37]{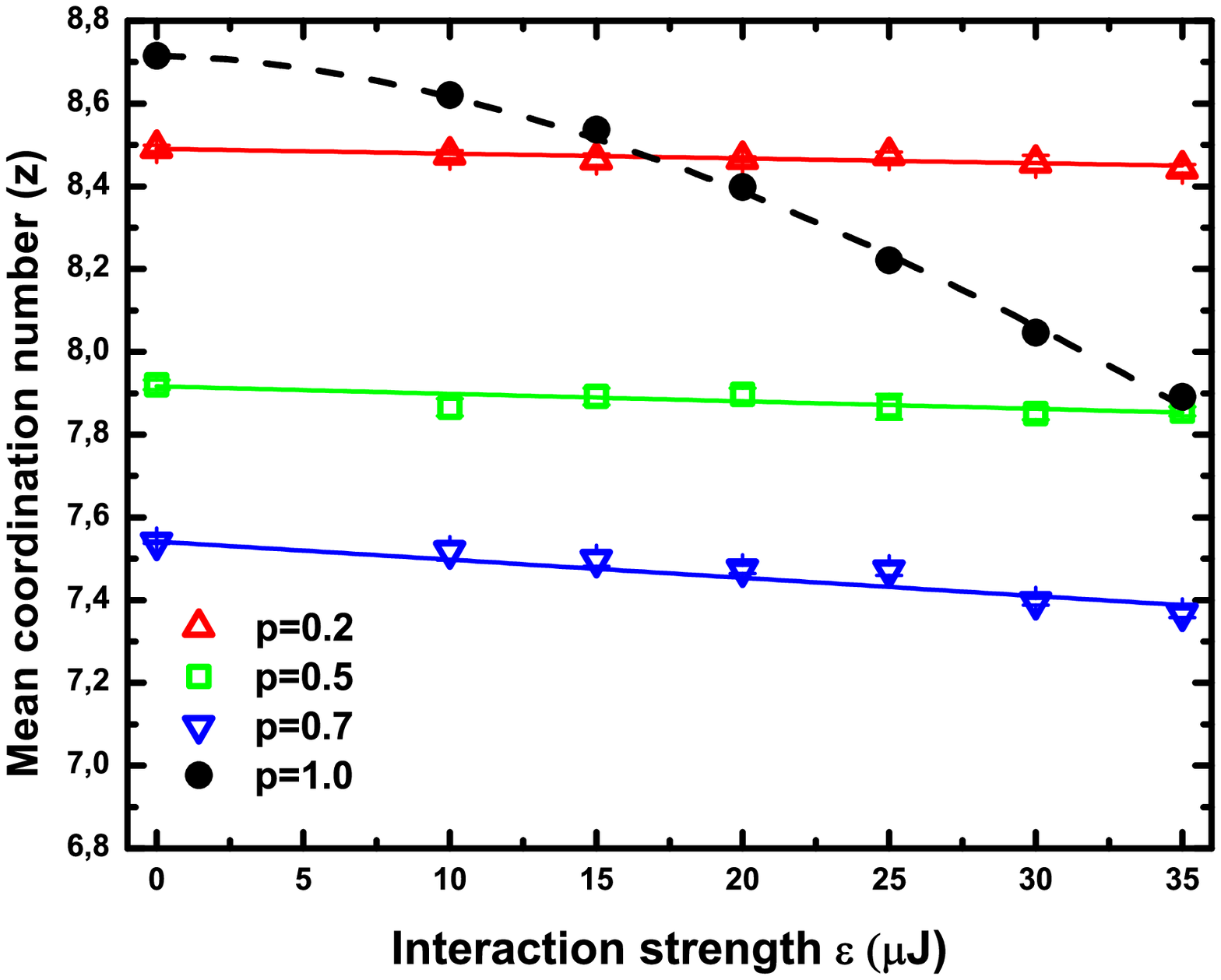}
 \caption{Plot of the  mean coordination number $z$ as a function of the interaction strengths $\varepsilon$ for all cases $p$ considered. Likewise, lines and error bars are as in Fig.~\ref{fig:04}.} \label{fig:05}
\end{minipage}
\end{figure*}

The packing densities $\phi$ for different $p$ cases, and considering several long-range interaction strengths $\varepsilon$, are shown as a function of time in Fig.~\ref{fig:03}. In this figure, the $\phi$ values are given at short time intervals of $1.0\, \mu$s up to 20 ms. At 20 ms, the ultimate $\phi$ values were obtained for each case. The initial packing densities were $0,22$ for the case $p=0.20$; $0,15$ for the case $p=0.50$; $0.10$ for the case $p=0.70$; and $0.26$ for the monodispersive case ($p=1.0$). The packing density minimum around 2.0 ms was due to the first particles' bouncing after hitting the bottom base of the box. In all cases, the ultimate $\phi$ values were below $\pi/\sqrt{18}\simeq 0.74$~\cite{zamponi2008}, which corresponds to closest-packing crystal structures, namely, face-centered cubic (fcc) and hexagonal close-packed (hcp) structures. As an overall behavior, one can see that packing density decreases with increasing $p$ value. This means that when one increases the relative number of large particles to small ones, it is seen an increase of the packing density in the formed structures. This is in accordance with other computational studies found in literature~\cite{clarke87,he99,jia12,Desmond2014}. A similar behavior was also found for the mean coordination number of the particles $z$. The mean coordination number $z$ is the number of neighboring particles that touch a given particle. A neighboring particle is found when the bond distance between two particles is equal the sum of their radii. For both $p=0.20$ and $p=0.50$ cases, the binary samples possess relatively higher $\phi$ values than monodispersive samples ($p=1.0$). These higher $\phi$ values for these cases can be attributed to the existence of a greater number of large particles in the granular material that either are wrapped around by smaller particles or create interstices that are filled by these. Furthermore, it is seen that for the cases $p=0.20$ and $p=0.50$, there have not been any meaningful influence of the long-range forces over the packing observables, namely $\phi$ and $z$. 
\begin{figure*}[t]
\centering
\begin{minipage} [t]{0.49\linewidth}
\centering
\includegraphics*[scale=0.37]{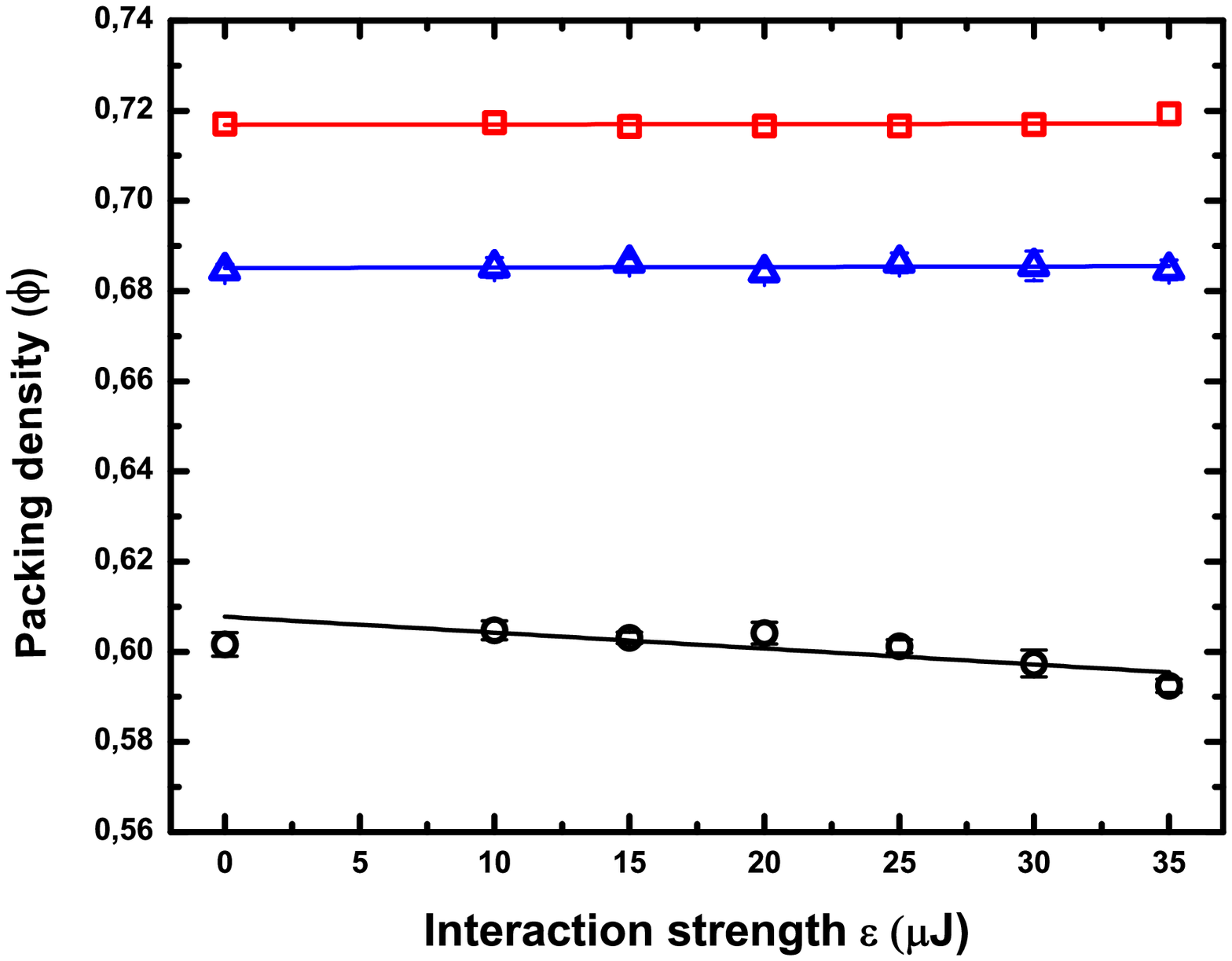}
 \caption{Plot of $\phi$ against $\varepsilon$ for the case $p=0.70$ considering different values of the size ratio $\lambda$. Black circles represent the case $\lambda=1/2$ ($a=4.0\, \mu$m, $b=2.0\, \mu$m), red squares represent the case $\lambda=1/3$ ($a=6.0\, \mu$m, $b=2.0\, \mu$m) and blue triangles represent the case $\lambda=1/4$ ($a=8.0\, \mu$m, $b=2.0\, \mu$m).  The straight lines are linear regressions to data.} \label{fig:06}
\end{minipage}\hfill
\begin{minipage}[t]{0.49\linewidth}
\centering
\includegraphics[scale=0.36]{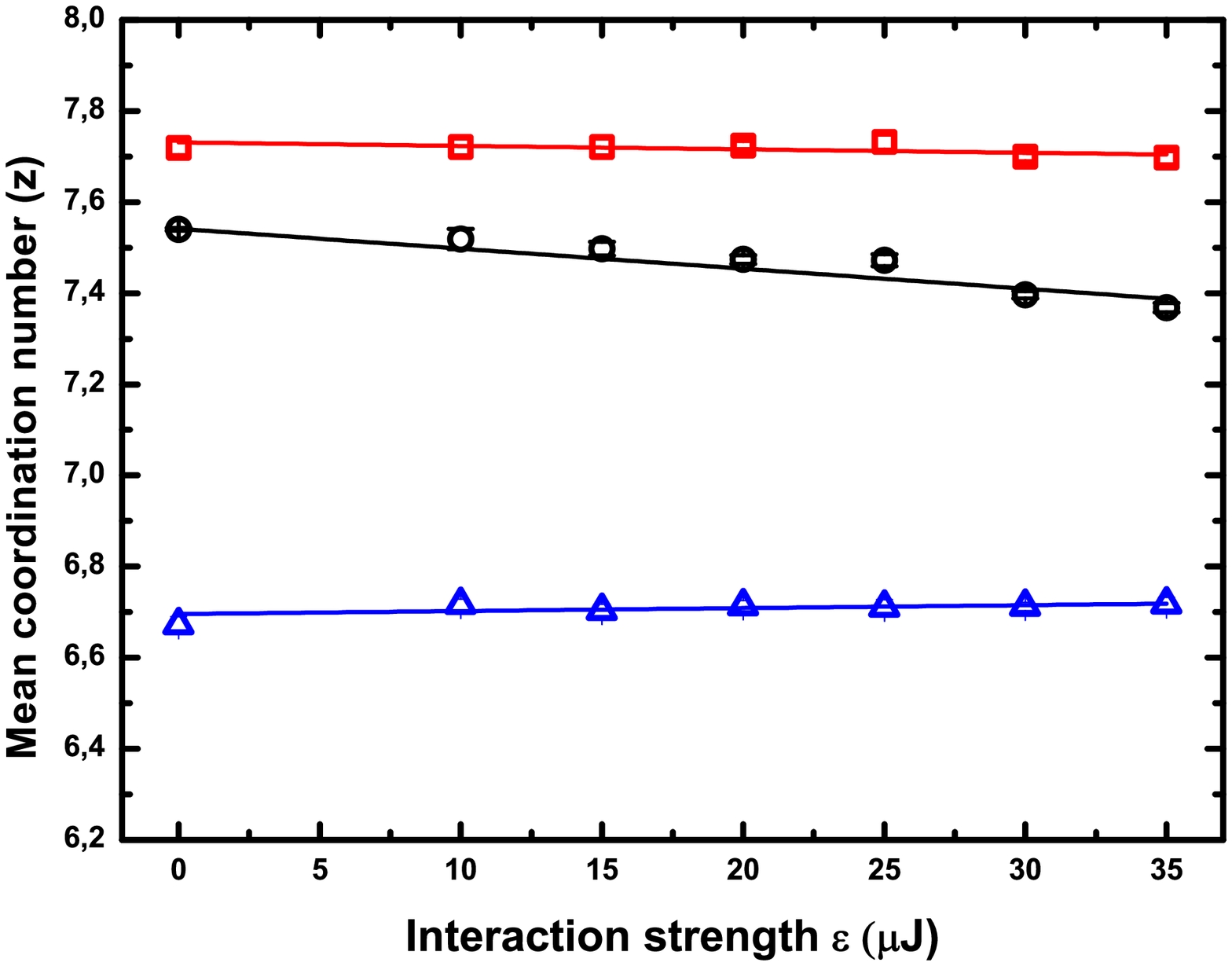}
 \caption{Plot of $z$ against $\varepsilon$ for the case $p=0.70$ considering different values of the size ratio $\lambda$. The symbols and lines are as in Fig.~\ref{fig:06}.} \label{fig:07}
\end{minipage}
\end{figure*}

On the other hand, meaningful effects of the long-range forces on these observables were only found for the cases $p>0.70$, especially for the case $p=1.0$. This behavior is made clearer by plotting the ultimate values of $\phi$ and $z$ versus $\varepsilon$, as shown in Figs.~\ref{fig:04} and \ref{fig:05} concerning some treated cases. Strikingly, for the cases $p=0.20$ and $p=0.50$, the slopes of the straight lines that fit the points are found to be practically zero ($\sim 10^{ - 5}$), revealing that the long-range forces are not effective to influence neither the packing dynamics (see Fig.~\ref{fig:03}) nor the ultimate $\phi$ and $z$ values. Even for the case $p=0.70$, such forces had very little influence on these processes. Figs.~\ref{fig:06} and \ref{fig:07} display the quantities $\phi$ and $z$ against $\varepsilon$ for the case $p=0.70$, considering different values of the size ratio $\lambda$. Only for samples with $\lambda=1/2$, it was found a weak $\varepsilon$ dependence. Whereas samples with other $\lambda$ values were found to be insensitive to the action of the long-range forces. Conversely, for the monodispersive case ($p=1.0$), the long-range forces had a meaningful influence on both dynamics and final packing quantities. For samples generated by distributions with high $p$ values, the $\phi$ values decreased with increasing interaction strength $\varepsilon$. This behavior is in accordance with experimental results obtained by Forsyth {\it et al}~\cite{forsyth2001} and with prior computational studies~\cite{yang2000,zou2008,gonzalez2014,handrey2018} for monosized particles. Different from liquids, particles in granular matters possess much more freedom to reorganize themselves by action of the long-range forces, and therefore, can more easily alter their structural properties when the magnitude of the long-range interaction strength changes.

Defining the largest relative amplitude for an observable $y\equiv \phi, z$ as 
\begin{equation}\label{eq:15}	
\delta \hat y  = \left(\frac{{y_{max} - y_{min}}}{{y_{max}}}\right) \times 100\%,
\end{equation}
where $y_{max}$ and $y_{min}$ are the maximum and minimum values of an given observable $y$, one can have some information about the effectiveness of the long-range forces in each case. From the simulation results, we obtained by taking $\lambda=1/2$: $\delta \hat \phi=1.66$\% and $\delta \hat z=2.25$\% for the case $p=0.70$; $\delta \hat \phi=3.51$\% and $\delta \hat z=3.41$\% for the case $p=0.80$; $\delta \hat \phi=13.21$\% and $\delta \hat z=8.66$\% for the case $p=0.90$; and $\delta \hat \phi=12.80$\% and $\delta \hat z=9.52$\% for the case $p=1.0$. For remaining cases, the largest relative amplitudes $\delta \hat y$ turned out to be less than $1.0$\%. 

In particular, for the case $p=1.0$, the data trend shown in Figs.~\ref{fig:04} and \ref{fig:05} follow a non-linear decrease with increasing $\varepsilon$ value. After attempting different prescription models to model these data, better fits seem to be achieved by using the following expression 
\begin{equation} \label{eq:16}
y = y_0 \exp ( - B \varepsilon ^\alpha ),
\end{equation}
where $\alpha$ and $B$ are fitting parameters. These parameters are: $\alpha=2.57$ and $B=1.51 \times 10^{-5}$ for $\phi$ (with a least reduced chi-squared $\chi_{\nu}^{2}=2.57$ and a goodness-of-fit factor $Q=3.6$\%); and $\alpha=1.75$ and $B=2.03 \times 10^{-4}$ for $z$ ($\chi_{\nu}^{2}=2.62$ and $Q=3.3$\%). Such behavior for $\phi$ and $z$ as the long-range force strength increases has also been corroborated by another computational study concerning different particle size distributions ~\cite{handrey2019}.

The RDF has been widely used to characterize random structures of spherical particles~\cite{yen91,jia12}, where it can be understood as the probability of finding one particle at a given distance from the center of a reference particle. The RDF is defined as
\begin{equation}\label{eq:17}
 g(r_{i})=\dfrac{n(r_{i})}{4\pi \, r_{i}^{2} \, \delta r_{i}\, \rho},
\end{equation}
where $n(r_{i})$ the number of particle centers within the $i$-th spherical shell of radius $r_{i}$ and thickness $\delta r_{i}$, and $\rho$ is the number of particles per volume. In the above equation, we set $\delta r_{i}=0.1 \, \mu$m, and regard a number of spherical shells $N_{r}=150$ for $g(r)$ computation.

 The reasons for calculating $g(r)$ for bidispersive packings are two-fold: first, we would like to know how $g(r)$ profiles change by considering different $p$ values and, secondly if such profiles are also altered when the long-range interaction strength $\varepsilon$ is increased. As we will see, the $g(r)$ profiles change for different $p$ values, and are sensitive only for large enough $\varepsilon$ values. 

\begin{figure*}[!t]
 \centering
 \includegraphics[scale=0.70]{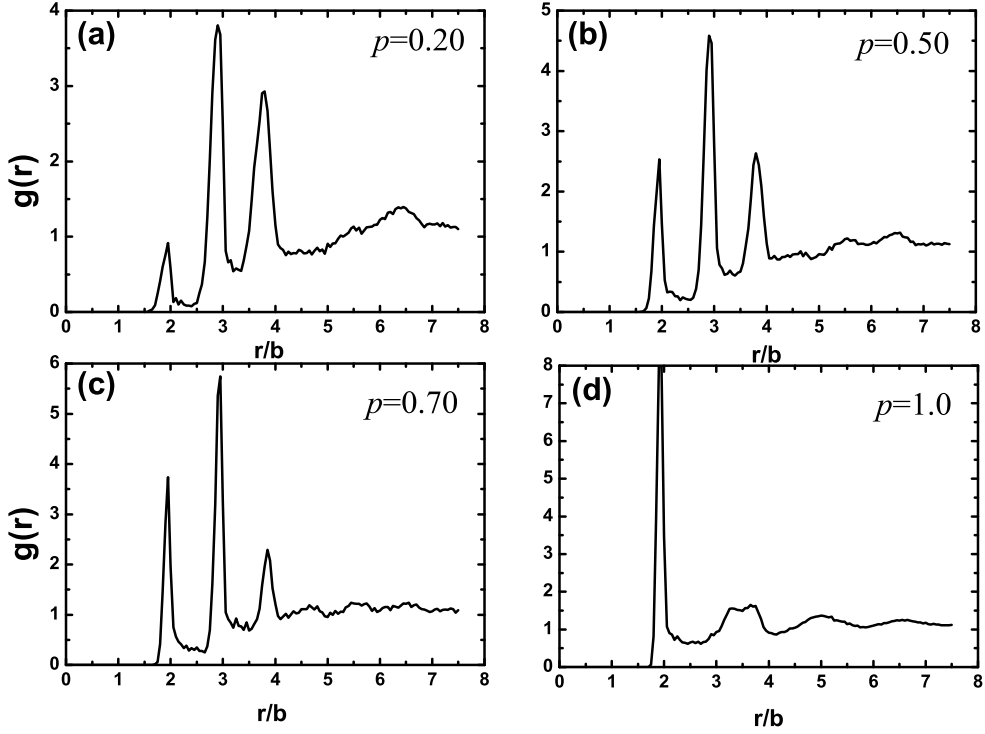}
 \caption{Typical RDF of the random packing structures formed by using different $p$ values when $\varepsilon=10.0\, \mu$J and size ratio $\lambda=1/2$ ($a=4.0\, \mu$m, $b=2.0\, \mu$m). The above panels display the cases: a) $p=0.20$, b) $p=0.50$, c) $p=0.70$ and d) $p=1.0$. The force ratios (Eq.~\ref{eq:14b}) are: $\zeta=0.04$ for $p=0.20$; $\zeta=0.06$ for $p=0.50$; $\zeta=0.09$ for $p=0.70$; and $\zeta=0.49$ for $p=1.0$.} \label{fig:08}
\end{figure*}

Fig.~\ref{fig:08} shows typical RDFs as a function of the radial distance for the random packing structures by considering different $p$ values when $\varepsilon=10.0\, \mu$J and $\lambda=1/2$ ($a=2.0 \,\mu$m, $b=4.0\,\mu$m). These curves were obtained by averaging the individual RDFs of all particles inside the bulk region of the formed particle aggregate. This bulk region is defined here as a smaller virtual box centered at the central point of the aggregate and having an offset distance of $15\;\mu$m from each actual wall of the confining box. From Fig.~\ref{fig:08}, one can observe three main peaks in RDF for the bidispersive cases ($p<1.0$). Such peaks are localized around the distances $4\, \mu$m, $6\, \mu$m and $8\, \mu$m. These three peak values correspond to three different contact types, namely, contact of type I: smaller particles to smaller particles; contact of type II: smaller particles to larger particles; and contact of type III: larger particles to larger particles. For the case $p=0.20$, contacts of type III are more frequently found; for the case $p=0.50$, contacts of type II are more frequently found; and for the case $p=0.7$, the contacts of type I happen more frequently. While for the case $p=1.0$, it is seen the well-known peaks localized at the distances $4\, \mu$m and close to $4\sqrt{3}\, \mu$m (split second peak), which corresponds to spheres arranged in pairs of coplanar equilateral triangles which share a common edge~\cite{finney1970}. 

\begin{figure*}[t]
\centering
\begin{minipage} [t]{0.49\linewidth}
\includegraphics*[scale=0.36]{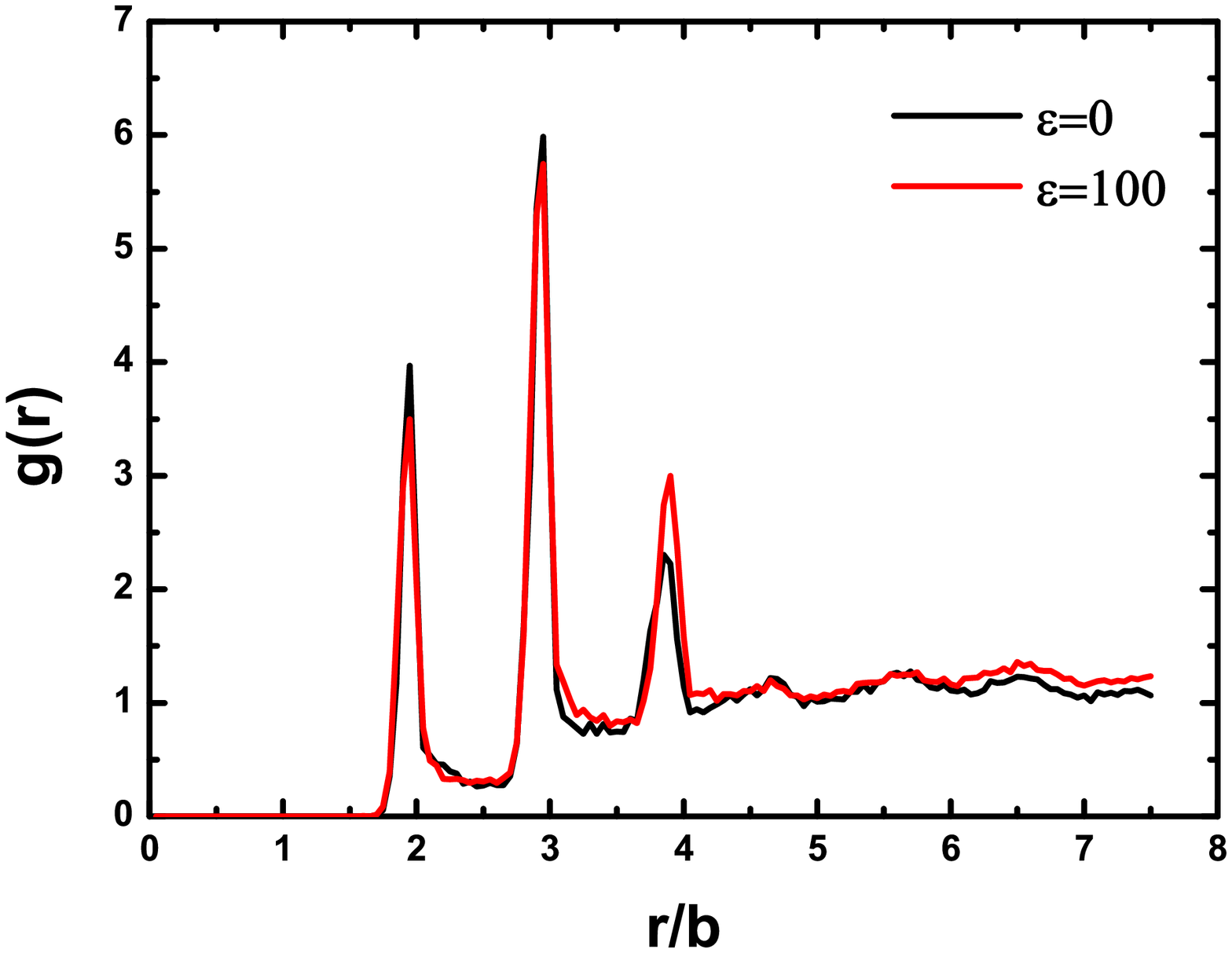}
\centering
 \caption{RDFs of the random packing structures formed for the case $p=0.70$, considering two different $\varepsilon$ value with a size ratio of $\lambda=1/2$ ($a=4.0\, \mu$m, $b=2.0\, \mu$m). The black line represents RDF when $\varepsilon=0\, \mu$J (absence of long-range interactions) and the red line represents RDF when $\varepsilon=100\, \mu$J (the maximum $\varepsilon$ value considered here). The force ratios (Eq.~\ref{eq:14b}) are: $\zeta=0.0$ for $\varepsilon=0\, \mu$J and $\zeta=0.82$ for $\varepsilon=100\, \mu$J.} \label{fig:09}
\end{minipage}\hfill
\begin{minipage}[t]{0.49\linewidth}
\includegraphics[scale=0.36]{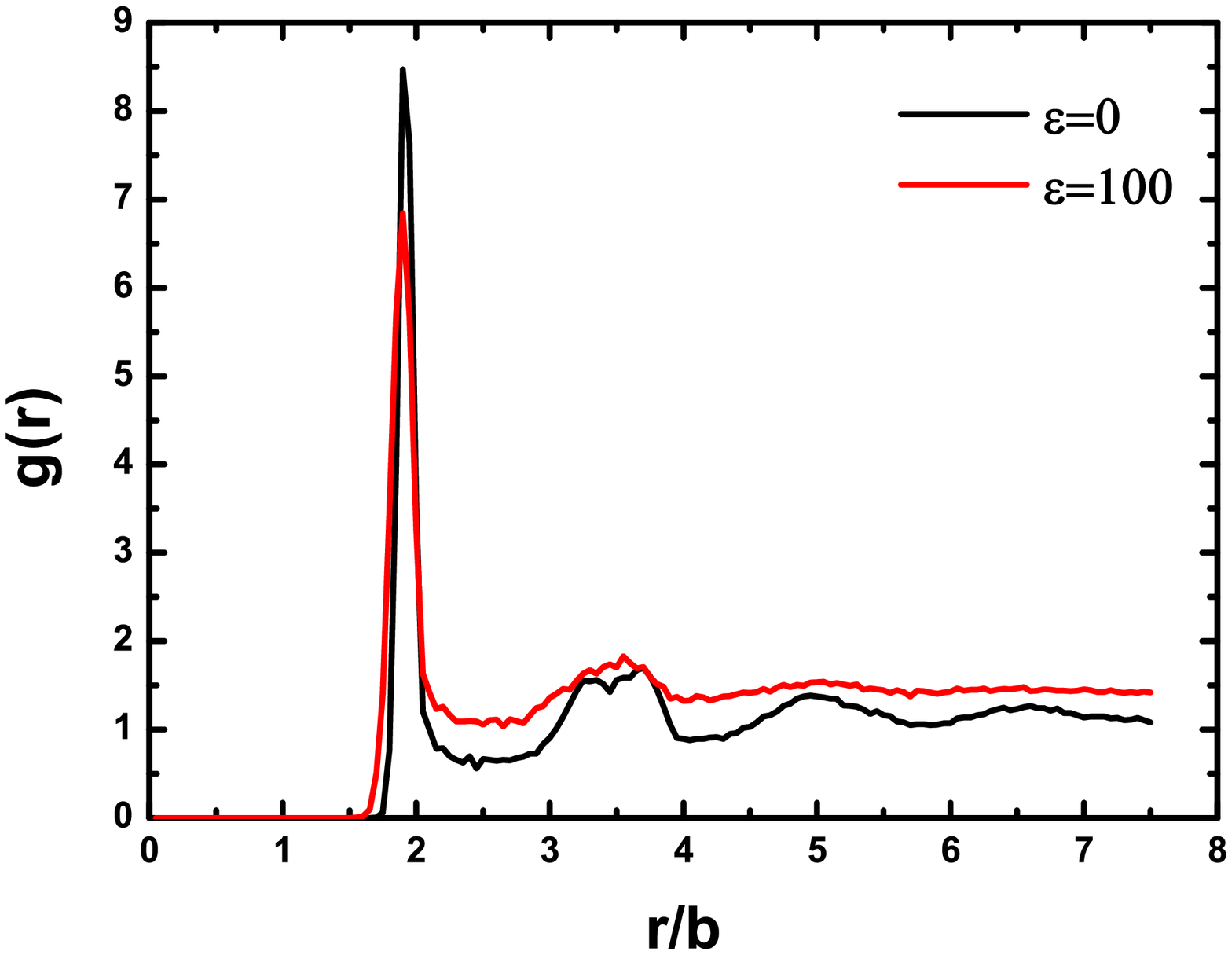}
\centering
 \caption{RDFs of the random packing structures formed for the case $p=1.0$, considering two different $\varepsilon$ value and particle radius equals to $2.0\,\mu$m. The lines are as in Fig.~\ref{fig:09}. The force ratios are: $\zeta=0.0$ for $\varepsilon=0\, \mu$J and $\zeta=4.53$ for $\varepsilon=100\, \mu$J.} \label{fig:10}
\end{minipage}
\end{figure*}

Once the samples generated in the present study through sedimentation process are rapidly brought to the random close packing (RCP) state, only long-range interaction strengths larger than those in the range $\varepsilon=0-35\, \mu$J, which have been considered here, can significantly change their corresponding $g(r)$ profiles. The Figs. ~\ref{fig:09} and ~\ref{fig:10} show this sharp difference in the RDFs for both outstanding cases $p=0.70$ and $p=1.0$, when a long-range interaction strength with two magnitude order ($\varepsilon=10^{2}\, \mu$J) is used during packing process in comparison to when $\varepsilon=0\, \mu$J, which means absence of long-range interactions. One can see that the general shape of the RDFs reflects the binary particle distribution for each case, but significant changes are found when taking large enough $\varepsilon$ values in the simulations. In particular, the main peaks are lowered due to persistent action of the strong long-range forces, suggesting that these forces can lead the particles to a less dense and more disordered state. For the case $p=1.0$, it is even easier to see that the peak values are clearly diminished by the action of the long-range forces in comparison to when such forces are absent.  In fact, the more large particles are present in the binary particle packing, the more attenuated are the effects of the long-range forces. Thus one could efficiently mitigate the effects of the long-range forces during packing processes by decreasing the population density $p$ of small particles, at least within the specific conditions assumed here. 

For further analysis, it is also desirable to understand how the long-range forces change the orientational order of the formed structures. To accomplish that, let us first define a local structural parameter sensitive to ordering as
\begin{equation}\label{eq:19}
\psi _{i,m}^l  = \frac{K}{{n_b }}\sum\limits_{j = 1}^{n_b} {Y_{lm} (\theta _{ij} ,\varphi _{ij} ),} 
\end{equation}
where $\psi _i ^l$ is a complex vector with $|m|\leq l$ components assigned to every particle $i$ in the system. In Eq.\ref{eq:19}, the sum runs over all $n_b$ nearest neighbors of the particle $i$ and $K$ is a normalization constant so that the complex inner product $\sum\limits_m^{} {\psi _{i,m}^l} \,\psi _{i,m}^{l*}  = 1$. For a given particle pair ($i$,$j$), $Y_{lm} (\theta _{ij} ,\varphi _{ij})$ is the spherical harmonic associated with the bond vector $\vec r_{ij}$ connecting this pair, being $\theta _{ij}$ and $\varphi _{ij}$ the according polar angles of this vector relative to some fixed coordination system. In order to check for both cubic symmetry (e.g., fcc and bcc structures) and icosahedral symmetry, we have taken $l=6$ (and $-6\leq m \leq 6$) in Eq.~\ref{eq:16}. Thus we deal with $N$ 13-element complex vectors for bond-ordering analysis.

Once the order parameter $\psi _i ^l$ has been defined, further information about the bond-orientational order can be obtained from the orientation pair correlation function given by
\begin{equation}\label{eq:20}	
G _{6} (\vec r) = \sum\limits_{m} {<\psi _{i,m}^{l = 6} (\vec 0)\,\psi _{j,m}^{l = 6} (\vec r)^* >}.
\end{equation}
   
In the above equation, the angular brackets indicate an average over all particles separated by $\vec r$ in the bulk. It is known that the ``bond" between any two particles $i$ and $j$ is recognized as crystal-like if $G _{6} (\vec r)>0.5$~\cite{steinhardt83,Desmond2009}. Thus we can check for crystallization, as well as changes in the ordering of the particles, due to the action of the long-range forces by computing $G _{6}$ as a function of the radial distance between particles. 

\begin{figure*}[!t]
\centering
\begin{minipage} [t]{0.49\linewidth}
\includegraphics*[scale=0.36]{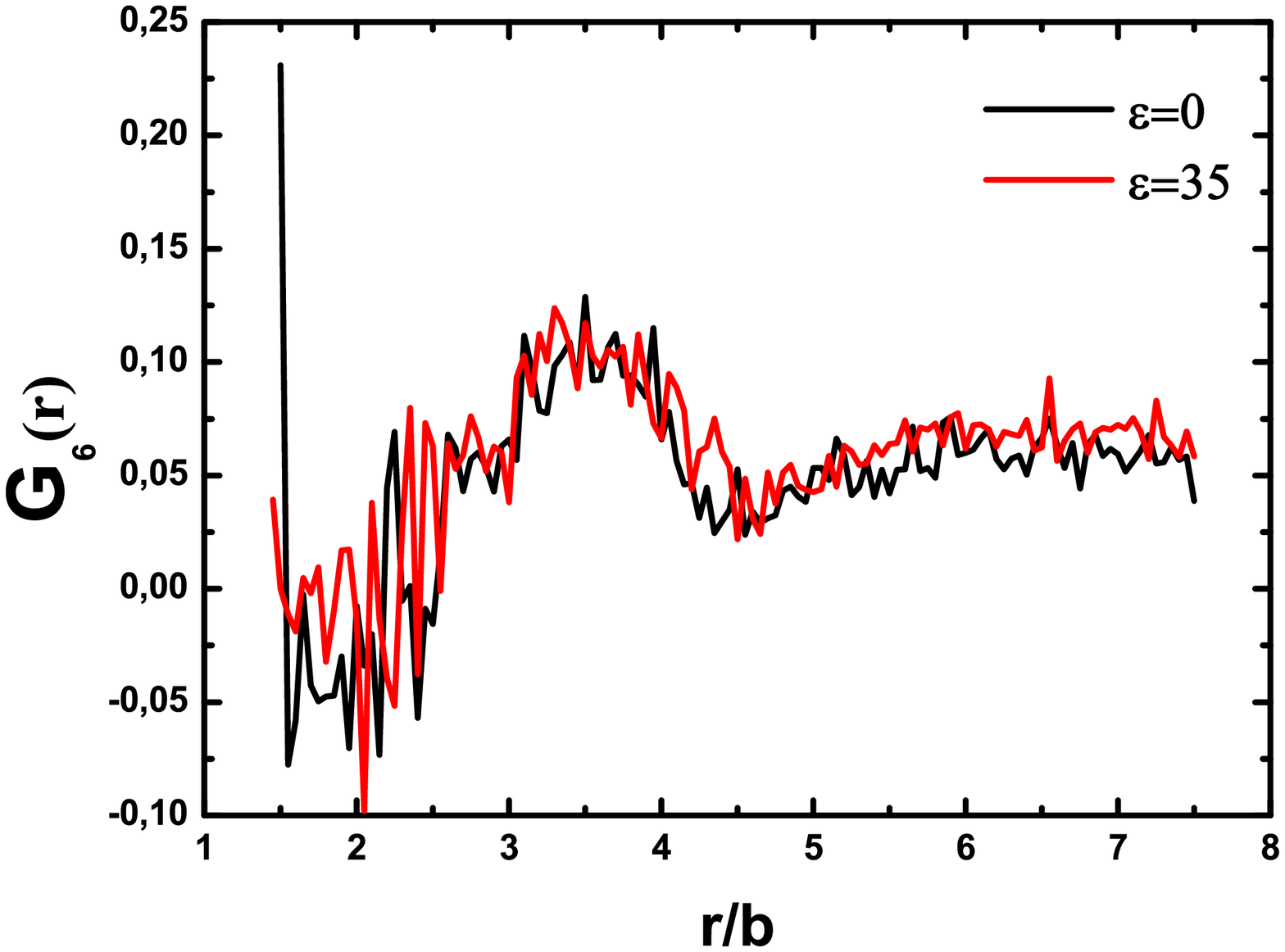}
\centering
 \caption{Orientation pair correlation function $G_{6}$ as a function of the radial distance for the case $p=0.70$, considering two different $\varepsilon$ value with a size ratio of $\lambda=1/2$ ($a=4.0\, \mu$m, $b=2.0\, \mu$m). The black line represents $G_{6}$ when $\varepsilon=0\, \mu$J  and the red line represents $G_{6}$ when $\varepsilon=35\, \mu$J. The force ratios (Eq.~\ref{eq:14b}) are: $\zeta=0.0$ for $\varepsilon=0\, \mu$J and $\zeta=0.31$ for $\varepsilon=100\, \mu$J.} \label{fig:11}
\end{minipage}\hfill
\begin{minipage}[t]{0.49\linewidth}
\includegraphics[scale=0.36]{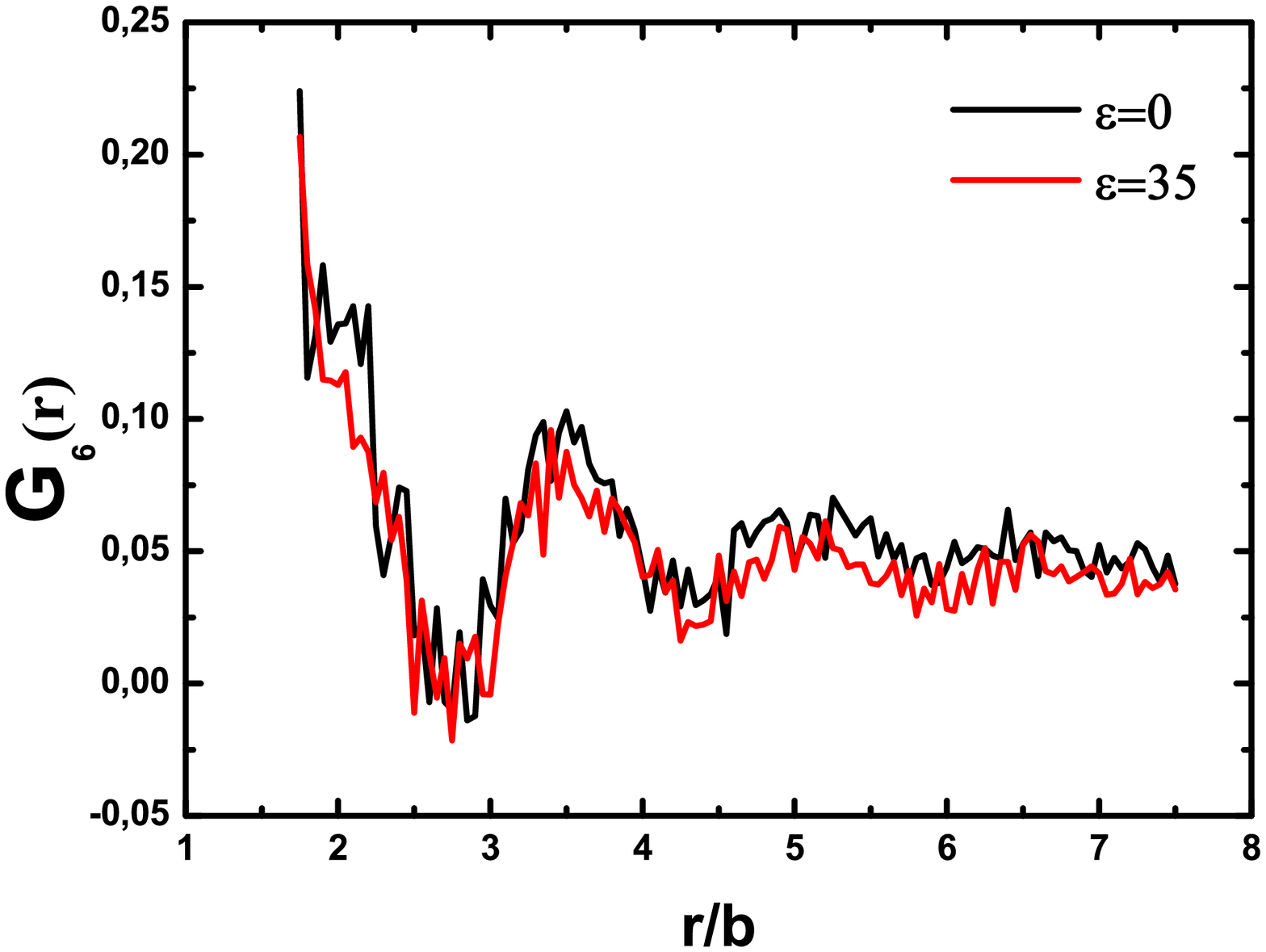}
\centering
 \caption{ Orientation pair correlation function $G_{6}$ as a function of the radial distance for the case $p=1.0$, considering two different $\varepsilon$ value and particle radius equals to $2.0\,\mu$m. The lines are as in Fig.~\ref{fig:11}. The force ratios are: $\zeta=0.0$ for $\varepsilon=0\, \mu$J and $\zeta=1.69$ for $\varepsilon=100\, \mu$J.} \label{fig:12}
\end{minipage}
\end{figure*}

Figs.~\ref{fig:11} and ~\ref{fig:12} display $G _{6}$ as a function of the radial distance, respectively, for the cases $p=0.70$ and $p=1.0$. We can observe that high values of $G_{6}$ are localized between the last two main RDF peaks for the case $p=0.70$ (see Fig.~\ref{fig:09}), and between first and second sub-peaks at the split second peak of RDF for the case $p=1.0$ (see Fig.~\ref{fig:10}). For the case $p=0.70$, $G_{6}$ converges to an asymptotic average value of $0.06$ when $\varepsilon=0\, \mu$J, and of $0.07$ when $\varepsilon=35\,\mu$J. While for the case $p=1.0$, $G_{6}$ converges to an asymptotic average value of $0.05$ when $\varepsilon=0\, \mu$J, and of $0.04$ when $\varepsilon=35\,\mu$J. It is interesting to observe the overall effect of the long-range forces over the bond-orientational order of the samples. For the monodispersive case, the action of these forces has, on average, reduced the bond-orientational order of the samples. In contrast, these forces have enhanced the bond-orientational order of the samples for the bidispersive case.   
 
 Finally, it is important to stress that the present results obtained through particle sedimentation mechanism may be different from those obtained by using other methods. By changing the protocol for generating such packings, one may obtain slightly different results. For instance, it is known that packings generated through collective rearrangement methods have given higher packing densities~\cite{clarke87,he99}.

\section{\label{sec:c} Conclusions}

In this study, MD simulations were performed to study the random close packing of spherical particles at micrometer scales. Both contact forces and long-range dispersive forces were taken into account in these simulations. Several cases of the binary size distribution were treated by changing different control parameters for constructing the samples, such as population density $p$, size ratio $\lambda$, and long-range interaction strength $\varepsilon$. The packing dynamics was studied by evaluating over time different physical observables, including the packing density, mean coordination number, and kinetic energy. Furthermore, RDF and orientation pair correlation function were calculated to characterize the particle structures that were formed over different values of the long-range interaction strength $\varepsilon$. It was found that packing dynamics is quite sensitive to both changes in $p$ and $\lambda$ values; nevertheless, being practically insensitive to changes in $\varepsilon$ values for most of the studied cases.

 Only for cases above $p=0.70$, there were effective influences of the long-range forces on packing processes. In particular, for the case $p=0.70$ with $\lambda=1/2$ was observed a weak influence of the long-range forces. While for the case $p=1.0$ was observed the strongest influence of these forces on packing processes. Both the packing density $\phi$ and mean coordination number $z$ gradually decayed as the $\varepsilon$ value increased in a different way for each case. For the case $p=0.70$, both $\phi$ and $z$ decreased linearly with increasing $\varepsilon$ value, while for the case $p=1.0$, these quantities decreased non-linearly with increasing $\varepsilon$ value.  

The general shape of the obtained RDFs reflected the binary distribution of the particles for each case, where it was practically unchanged by the long-range interaction forces for most treated cases. However, for the case $p=1.0$ was clearer to notice the influence of these forces on the rearrangement of the particles during the formation of the final packing structure. The RDF peaks were slightly diminished by the action of the long-range forces in comparison to when such forces were absent during the packing process. Furthermore, it was observed that the particle bond-orientational order is quite sensitive to the action of the long-range forces. These results may indicate that long-range forces hinder these processes by creating both additional voids and local particle clusters in the bulk of the formed structures.

In the present study was shown that the packing dynamics of particles with binary size distribution is quite resilient with respect to increases of the long-range forces, irrespective of the different $\lambda$ values considered. Striking exceptions are the cases with high $p$ values, especially $p=1.0$, where long-range forces were able to strongly influence the packing processes, particularly affecting quantities as packing density, mean coordination number and orientation pair correlation function. This is important because of its potential application to the development and fabrication of novel materials such as in sintering of both metallic powders and ceramics, and in modeling the atomic structure of amorphous metals composed of similar-sized atoms as in binary alloys.

\section{Acknowledgements}
We wish to thank UFERSA for computational support.




\bibliographystyle{model1a-num-names}

\end{document}